\documentclass{article}

\usepackage[english]{babel}
\usepackage{ntheorem}
\usepackage{wrapfig}
\usepackage{graphicx}
\usepackage{caption}
\usepackage{subcaption}
\usepackage[export]{adjustbox}
\usepackage[letterpaper,top=2cm,bottom=2cm,left=3cm,right=3cm,marginparwidth=1.75cm]{geometry}
\usepackage[parfill]{parskip}

\usepackage{amsmath}
\usepackage{graphicx}
\usepackage[colorlinks=true, allcolors=blue]{hyperref}

\usepackage{comment}

\title{Strategic Communication and Deliberation on Climate Change of different Actor Groups using Twitter}
\author{Julian Dehne and Valentin Gold}

\begin{document}

\frenchspacing

\clubpenalty10000
\widowpenalty10000
\displaywidowpenalty=10000

\hyphenpenalty=9000
\exhyphenpenalty=8000

\maketitle

\begin{abstract}
Strategic communication in Twitter is compared between different actor groups with regard to the topic of climate change. The main hypothesis is that different actor groups will be more or less central in the reply-trees depending on their strategic interests based on their profession or organizational affiliation. 
\end{abstract}

\section{Introduction}

Following the SARS-CoV-2 epidemic, a lot of attention has been drawn to the sinking effectiveness of institutionalized communication in liberal democracies in the case of issues that have a complex scientific background. In parts this can be attributed to a fractured public forum and reduced trust in the credibility of experts. But there is a need to measure this perceived trend or it should be disregarded as speculation. In order to formalize a relationship that describes a trend for this institutionalised communication, the sender of the communication needs to be defined more narrowly and the expected effect, too. For this purpose the concept of strategic communication is introduced. Strategic communication \cite[1]{hallahan_defining_2007} needs to be conceived not only regarding its penetration but also on how stable the newly created information bubbles are. Another factor is the complexity of the scientific underpinnings of the message that needed to be heard. For this reason the focus will be on strategic \textsl{scientific} communication of the politically polarizing topic (such as vaccination, migration, climate change \ldots) that are complex but need a functioning discursive space in order to allow for social change. In order to narrow down the actors and relevant language, climate change was picked as a case study.

After conceptualizing the starting point as strategic communication, the range of the effect needs to be mapped out and reduced to an observable subset of the public sphere. In order to access the quantifiable part of the public communication sphere, social media text data is used (for example from Twitter, Reddit). In the end, the most important question is what constitutes successful strategic communication in the context specified (social media, polarizing scientific complex topics, institutionalized actors) if this phenomenon is to be measured at all.

One important part of this question is how the success of strategic communication should be conceived. \cite[48]{strategic_turn} argues that the strategic turn has reduced participation to a means and not a goal in itself. For example, a politician would use scientific facts rhetorically, in order to achieve some hidden agenda. Whether intended or not it will be assumed that the actor's strategic interest lies in some motivation to raising awareness of scientific facts, motivate social change according to these facts, and stimulate discourse in order to legitimate decisions based on them.

\section{Epistemic Bubbles and Strategic Scientific Communication}


Idealist political philosophy proposes that every opinion must have access to and be evaluated rationally by a power-free discourse \cite{christoph_lumer_habermas_1997}. Even pragmatic or rational choice approaches assume a minimum of extended rationality in the pursue of the player’s interests. However, both contraptions seem out of date considering the rise of echo chambers and epistemic bubbles \cite{nguyen_echo_2020}.
\begin{quote}
    An epistemic bubble is a social epistemic structure in which other relevant voices have been left out, perhaps accidentally. An echo chamber is a social epistemic structure from which other relevant voices have been actively excluded and discredited~\cite{nguyen_echo_2020}.
\end{quote}

Although Nguyen makes the valid point that echo chambers should be treated as a different phenomenon, both can be treated as roadblocks in strategic communication. Epistemic bubbles also include filter bubbles~\cite{garrett_echo_2009} \cite{flaxman_filter_2016}~\cite{bright_explaining_2017}, \cite{filter_bubbles}, where automated personalization leaves digital citizens stranded on an island of their own beliefs. More generally, epistemic bubbles are the limits even an approximately informed \cite{brita_ytre-arne_approximately_2018} citizen has when it wants to learn about public events. Conspiracy theories are particularly problematic from a governmental standpoint~\cite{keeley_conspiracy_1999}~\cite{romer_conspiracy_2020}. These are echo chambers that repeat beliefs that are not true according to the institutional position. These bubbles can be treated as opposites to peer-reviewed expert communication.

The concept of a bubble has a psychological, a sociological, and a sociometric model attached. From the psychological perspective, there are natural limits to information processing in the human mind. These limits are adequately modeled by the cognitive load theory \cite{sweller_cognitive_2011}. The more fractured the public space becomes, the harder it is to follow the discourse, even if the motivation was high. The sociological perspective is called "'groupstrapping"' \cite{boyd_epistemically_2019}. Boyd extends the communication model from Nguyen and adds group effects into the equation. Finally, the sociometric definition defines the bubbles as social network structures \cite{scott_social_1988}. The communication flows are paths along the edges. The vertices are the manifestations of the information in communicative action. The edges are the people that communicate. Where the discourse perspective makes it possible to connect to rhetorical studies and audience use, the network model allows for concepts from computer science to spill over, which is necessary to study the effects of the algorithms involved in personalization and the spread of news.



In order to measure the effectiveness of strategic communication, the classical approach addresses the effect the particular expert communication had on the public opinion. For instance, if a politician spoke out pro vaccination
the effectiveness would be measured in terms of how the public opinion on vaccinations has changed or how many people got vaccinated. However, this kind of model creates a black box around the area where the strategic communication attempt
was filtered: the social media. Although social media platforms do not bare the sole responsibility for shaping the perception of the communicated information nor are they the only social force that influences the epistemic bubbles they are of interest as they represent a view on the communication process that can be accessed directly in terms of data acquisition without creating an observer bias.

This way, the effect of strategic communication can be operationalized as the size of the created epistemic bubble \cite{nguyen_echo_2020} and its characteristics.
The latter will be analyzed from two perspectives. First, it will be (re-)constructed as a conjunctive room of experience \cite{kleemann_interpretative_2013} to generate a socio-genetic typology of the bubbles. Second, a bubble will be characterized as a social network structure \cite{scott_social_1988}. Here linguistics and natural language programming are applied.

However, the qualities of \textbf{group communication structures} in social media do not directly transfer to a measure of political strategic communication. Strategic communication can be viewed as successful in two dimensions. First, reaching a high percentage of citizens and second, stimulating meaningful and productive discourses that legitimate political decision making in democracies. Here, deliberation theory can be applied in order to quantify the effects in terms of improving discourses and transparent penetration of the public sphere \cite{chambers_deliberative_2003, thompson_deliberative_2008} .

Another aspect is the question of the content of the message send by the strategic communication act. \cite[2]{wenninger_social_2017} discuss several social changes that influence the communication of facts that are based on scientific research:

\begin{enumerate}
    \item the possibility for scientists to communicate directly to the public
    \item the competition of non-founded information
    \item the influence of the social media platforms and their algorithm
    \item the competition for trust and simplicity
\end{enumerate}

Regarding the topic climate change \cite[6]{fownes_twitter_2018} provides a comprehensive view on the different angles the connection between climate change and Twitter have been investigated: In general, Politicians' use of tweets is less investigated. ~\cite[1617]{golbeck_twitter_2010} sees signs that arguing is not one of the normal uses of Twitter for U.S. Congress members (presenting information is the strongest). The general picture seems to be that politicians use of Twitter both in qualitative as well as quantitative studies is focussed on transparancy, mini-press releases, outreach and sharing information \cite{aharony_twitter_2012, lamarre_tweeting_2013, silva_politicians_2022}. 

A clear quantitative measure of a certain strategic communication group is the number of actual replies in a conversation thread. Here only 3.7 \% of tweets counted of members of offices are replies \cite{mergel_connecting_2012}. This showcases that the strategic intentions and use of social media differ between groups of actors. This motivates the goal of this paper: mapping out the main actors on strategic scientific communication in relevant fields and analyzing the different ways they approach deliberation on social media. 

Assuming that these results can be generalized to other countries (i.e. Germany) and politicians that are concerned with the policy in question (climate change) the pattern of low engagement in discussions and a high ration of broadcast communication can be assumed. 

\newtheorem{hyp}{Hypothesis} 
\begin{hyp}
\label{hyp:outreach}
Politicians have a higher focus on outreach than other groups. 
\end{hyp}

\begin{hyp}
\label{hyp:sharing}
Scientists have a higher focus on information sharing than other groups.
\end{hyp}

\begin{hyp}
\label{hyp:debate}
Activists have a higher focus on debating issues than other groups.  
\end{hyp}

\begin{hyp}
\label{hyp:journalists}
Journalists' communication style includes debating issues but also information sharing and outreach.  
\end{hyp}

\begin{hyp}
\label{hyp:GO}
Governmental organisations behave similar to politicians in their communication style. 
\end{hyp}

The actor groups are defined by the sample drawn which included categorizing the different accounts. Further than that outreach, debating issues and information sharing need to be measured in order to test the hypotheses.

Outreach can be approximated using the out-degree of the nodes in the reply-tree, the root dominance\footnote{The Engagement API from Twitter is currently Business Only which prevents using views as another measurement}. Of course, the outdegree assumes that a high number of replies corresponds with the intention of achieving the latter. Posting original messages rather than entering an existing conversation does not proof the intend of outreach but it may be seen as a strong indicator. Furthermore, the intention of broadcasting for the purpose of outreach corresponds negatively with engagement in debate (or with low number of posts). 

Debating issues can be measured much more directly: high engagement is assumed to correspond to high author vision\footnote{Defined as the aggregated likelihood of an author having seen previous posts in a conversation. This measure was developed and published by the authors and is available on arXiv.org: Julian Dehne and Valentin Gold (2023), \textsl{Consistent, Central and Comprehensive Participation} on Social Media}, a high overall centrality of the posts in the author graph and a high quantity of posts.

Information sharing can be measured by the comparatively longer texts, the presence of links and a higher depth of the conversation with a lower outdegree of the informational post. Longer texts and the presence of links make sense intuitively. Having fewer replies but longer discussions differentiates the intention of sharing information that lead to new discussions versus broadcasting information for the sake of outreach. 

\pagebreak
\section{Study Design}

\subsection{Sampling}

The sample of this paper consists of 182 twitter users in total, separated into the groups politicians, activists, scientists, governmental organisations, non-governmental organisations and journalists. 
There are 24 politicians in the sample, 9 of which are tweeting in English and 15 in German. Most of the German users are part of the Green Party.
Furthermore there are 12 English speaking and 13 German speaking users that are considered climate activists. They are usually connected to Fridays For Future or other climate activist groups such as Ende Gelaende or Extinction Rebellion and claim to be climate activists in their twitter bio.

This paper also looks at 34 scientists, 15 of which are English speaking and 19 German. These scientists usually have a P.h.d. in Meteorology, Biology or other natural sciences.

The sample also includes 29 governmental organisations, 5 of which are part of the German government and 24 are tweeting in English. Most of the latter are part of International Unions such as the UN or the EU, or part of English speaking governments. 
45 of the users in our sample are considered non-governmental organisations. In the context of the discussions of climate change, climate activist groups such as Fridays For Future, Extinction Rebellion, Ende Geleande and Greenpeace are also considered NGOs. In fact, Fridays For Future's local groups take up most of the NGO-Sample. 37 of the NGOs are tweeting in German, the rest in English. The Sample for NGOs also includes informational twitter accounts that regularly tweet about climate change, such as 'taz Klima', an account run by the German newspaper 'taz'.

The last group in the sample are journalists. This paper looks at 25 different journalists, 10 of which are tweeting in English and 15 are tweeting in German. They are writing for different reliable news papers and most of them specialize in topics about climate change. 

\begin{samepage}

The following restrictions were in place during download: 

\begin{itemize}
    \item The root post, the post at the beginning of the downloaded conversation, was written by one of the climate authors. 
    \item The conversation is longer than 5 posts.
    \item The root post has to contain a word usually used in the context of the discussion of climate change
    \item The root author's profile is public
\end{itemize}
\end{samepage}

These restrictions are meant to rule out those conversations that are too short to analyse and find conversations that revolve around topics of climate change. 14 authors (8 organisations, 3 journalists, 2 activists and one scientist) are excluded from the sample because none of their tweets fit the criteria. 8234 conversations containing around 1,8 million tweets are analysed in the following steps. Around 23000 of the tweets were written by climate change authors. 

\subsection{Conversational Properties}

Before explaining the approach taken other more common alternatives need to be discarded: studying the conversational style with a qualitative approach leads to a detailed and succinct picture of conversational practices. For instance \cite{aharony_twitter_2012} looks at three political leaders in comparison. However, for a comparison of groups of actors this creates to big of a workload. Another typical way to analyze intention and patterns of social media engagement is natural language processing. Although there most certainly will be markers of outreach or raising awareness within the language used the model would need training and would only be applicable to a specific platform (Twitter) and a general field pertaining to the topic in question. Conversely, the reply-tree is a  platform-independent unit of analysis. It is also agnostic towards the topic or writing styles involved. 

The reply-tree itself can indicate what kind of discussion is at hand. For instance, a discussion with a high depth (length of the longest reply-chain) more likely resembles an offline deliberation than a mushroom structure where there is one original post and many replies to this root post. Another metric is the root author dominance which defines the probability that any post of reply tree is authored by the root author. More often than not (and more true in Reddit than it is in Twitter) the root author dominates the discussion which leads to a reduced deliberative quality.

Using a combination of the reply-tree structure and the knowledge which author has written which post one can predict whether or not an author has seen a lot of the replies or not. The simple model uses to two assumptions mentioned previously: an author has seen a post with a probability depending on its distance from the root post combined with the distance of the next answer in the reply tree of the same author.
\bigskip

\begin{equation}
    \zeta := P(SEEN\mid(V_{j},V_{i}))) = \sum(1/2)^{(\mid path(V_{j},V_{i}) \mid-1)} 
    \label{eq:pathdistance}
\end{equation}

\bigskip
The equation \ref{eq:pathdistance} reads as: the probability zeta of having seen node j is the average sum of the decay function of the path length between all the nodes i written by the given author and the node j. For example if the author has only written one subsequent answer to a post and this answer has a path distance of two replies to the post j than the probability of having seen j for the author would be $1/2^{(2-1)}=0.5$. If the path distance was 1 for a direct reply the exponent would be 0 and the probability 1. This measure is computed as an average for all existing path between node j and nodes i of the given author. Analogously, the root distance can be defined as

\bigskip
\begin{equation}
    \vartheta := P(SEEN\mid(V_{j},V_{i}))) = \sum(1/4)^{(\mid path(root,V_{j}) \mid-1)} 
    \label{eq:rootdistance}
\end{equation}

\bigskip
\begin{equation}
    \Rightarrow P(SEEN) = \zeta \cup \vartheta
\end{equation}

\bigskip
This measure models the distance to the root post as more relevant than the position of the reply. This is due to the dominant visual position of the original post in most platforms.

By using the authors as the nodes and one author having replied to another as directed edges one can derive a directed graph from the reply tree. Although this graph ignores the intensity of user interactions it represents the centrality of users within the conversation adequately. In social network theory there are three standard measures of centrality: in-betweenness centrality, closeness-centrality and Katz-centrality. In the context of this paper the centrality together with the number of posts can be used to indicate engagement and participation in the conversation. The lack thereof together with fewer posts can indicate information sharing behaviour as providing information in many cases may lead the thread to run dry as it does not stimulate an emotional response.

\section{Results}

 Figure \ref{fig:min-max} shows the mean of the branching factor, depth, centrality, baseline vision and root dominance for each author group. In order to compare all these parameters we used min-max normalization and used the mean of the normalized variables.
 There are a few indicators that hypothesis \ref{hyp:outreach} is true, that politicians do have a higher focus on outreach than other groups. The number of posts by the root author in each conversation is relatively low. However in relation to the number of all posts in a conversation (the 'root dominance') the number of posts by root authors who are politicians is neither particularly high nor low. These calculations show ambiguous results. The total count of tweets in conversations that were started by politicians is also relatively high: conversations by politicians contain an average of around 200 tweets. This supports the hypothesis \ref{hyp:outreach} because it suggests that politicians tweets spark interest.

 Although scientists are not the biggest group in the sample, almost one third of the downloaded tweets were written by scientists. This could mean that scientists are more active than other groups or tweet more frequently. This supports hypothesis \ref{hyp:sharing}, that scientists are focused on sharing information. Around 50\% of the tweets by scientists contain links which is an indicator that almost half of the downloaded tweets contain further information on the shared topic. However around 60\%  of the tweets by governmental as well as non-governmental organisations contain links. As figure 1 shows, the baseline vision of scientists is low, in comparison to other groups. This indicates that scientists are not particularly interested in the conversation following their tweets and are therefore not interested in discussing topics. Thus, the calculations do not show clear results for hypothesis \ref{hyp:sharing}. 
 
\begin{figure}[ht!]
\caption{Mean of Total counts of Tweets in each conversation}
\centering
\includegraphics[scale=0.7]{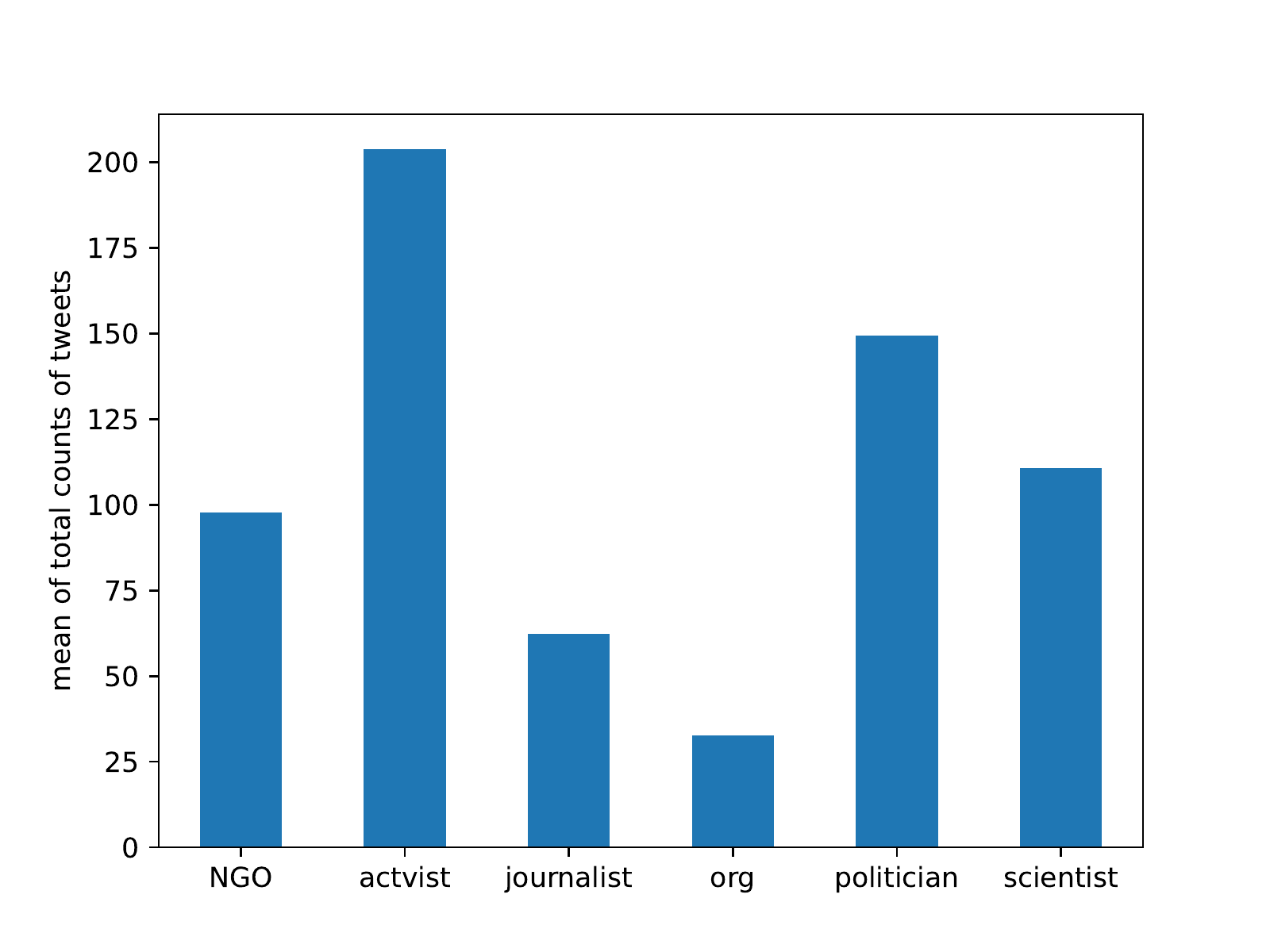}
\label{fig:total-counts}
\end{figure}

 Around 63\% of the downloaded tweets that were written by activists mention other twitter users which supports the hypothesis \ref{hyp:debate} that activists are focused on debating issues with others. Only around 35\% of these tweets contain links. One could conclude  that activists are not focused on sharing information. The tweets by the activists in our sample have a much higher number of replies than the other groups. On average there are 800 tweets in the conversations by activists. This might be due to outliers such as conversations by Greta Thunberg or Luisa Neubauer which usually have a very high engagement. These outliers were excluded from the sample for the visualisation in figure \ref{fig:total-counts}. Even without the outliers, activists have by far the highest reply count. As shown in figure 1, tweets by activists also have a high branching factor in comparison to other author groups. This shows that activists generally have a high engagement on their tweets. We conclude that activists are in fact more focused on debating issues than other groups and that the results support hypothesis \ref{hyp:debate}.

\begin{figure}[ht!]
\caption{Plot of min/max-normalized variables of participation of different strategic actor groups}
\includegraphics[width=1.3\textwidth,center]{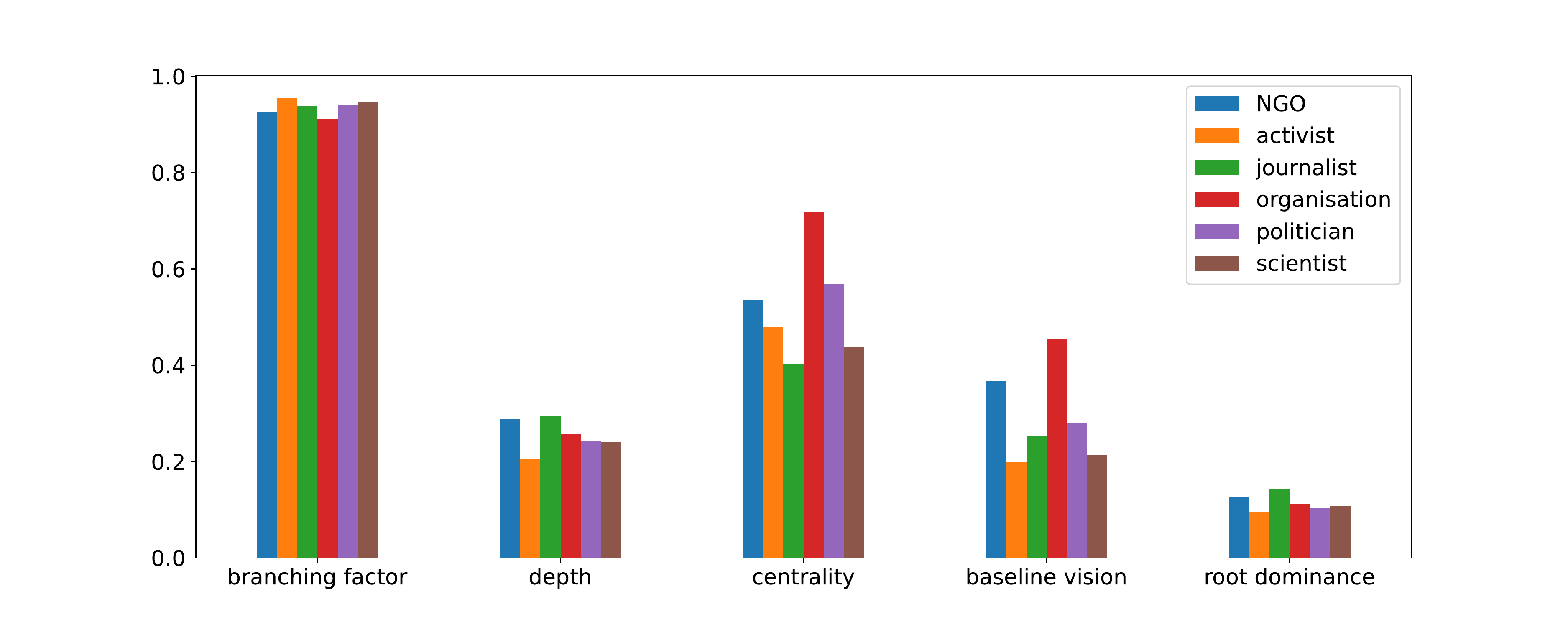}
\label{fig:min-max}
\end{figure}

 Only around 40\% of the tweets by journalists contain links, which might indicate that they are less focused on sharing information than in hypothesis \ref{hyp:journalists} presumed. The indicators that point to a debating conversational style are not found in the calculated variables for journalist. They are the group with the lowest baseline vision and the lowest centrality of all groups. Furthermore  figure \ref{fig:total-counts} shows that journalists are the group with the highest root dominance. This indicates that they are not focused on outreach and more on debating issues since a high root dominance indicates that the root author is involved in the conversation that they started.
 
 Tweets by governmental organizations have an average of 250 characters, whereas tweets by politicians have an average of 220 characters.~61\% of the tweets by politicians contain links while 51\% of the tweets by politicians do. Furthermore only 41\% of tweets by politicians mention other twitter users, while 61\% of the tweets of governmental organisations do. Other than expected in hypothesis \ref{hyp:GO} governmental organisations and politicians do not behave similar in their communication style. Other indicators also point to this statement. Organisations have a much higher root dominance and author baseline vision than politicians do. In general the engagement in conversations by organisations is by far the lowest. On average they have the lowest amount of replies. The calculations show that politician and governmental organisations on average have very different results for almost every calculated variable, except for the number of posts by the root author in each conversation, which is relatively low for both groups. This supports the hypothesis that both groups are more focused on outreach.


\section{Concluding Remarks}

In conclusion, the study reveals that politicians' communication style aligns with outreach, activists focus on debating issues, and journalists have a more debating-oriented style. However, the results challenge the hypotheses regarding scientists' information sharing focus and the assumed similarity between governmental organizations and politicians' communication style. These findings provide insights into the communication patterns of different author groups on Twitter.

The paper also highlights the use of the the more comprehensive concept of actor involvement in social media drawing from strategic communication research and methods from social network analysis. Further research is required to investigate how these measures could be validated by non-supervised methods. For instances, if author vision and centrality would cluster actor groups reliably, these measures could be used to identify hitherto unknown actor groups and map out epistemic bubbles. 

Another application of the methodology would be to analyze epistemic bubbles directly and compare the rhetoric of the leaders to the one of the outliers. In this case, the sampling would have to be focused on finding discussions that belong to a clearly defined bubble rather than using the actors as representative for their guild.



\bibliographystyle{alpha}
\bibliography{literatur_v1, mod_norm_theory_lit, ClimateChangeTwitter}

\newcommand{\etalchar}[1]{$^{#1}$}
\begin{thebibliography}{ZBTM{\etalchar{+}}16}

\bibitem[Aha12]{aharony_twitter_2012}
Noa Aharony.
\newblock Twitter use by three political leaders: an exploratory analysis.
\newblock {\em Online Information Review}, 36(4):587--603, January 2012.
\newblock Publisher: Emerald Group Publishing Limited.

\bibitem[Boy19]{boyd_epistemically_2019}
Kenneth Boyd.
\newblock Epistemically {Pernicious} {Groups} and the {Groupstrapping}
  {Problem}.
\newblock {\em Social Epistemology}, 33(1):61--73, January 2019.

\bibitem[Bri16]{bright_explaining_2017}
Jonathan Bright.
\newblock Explaining the emergence of echo chambers on social media: the role
  of ideology and extremism.
\newblock {\em CoRR}, abs/1609.05003, 2016.

\bibitem[Cha03]{chambers_deliberative_2003}
Simone Chambers.
\newblock Deliberative democratic theory.
\newblock {\em Annual Review of Political Science}, 6(1):307--326, January
  2003.

\bibitem[FGR16]{flaxman_filter_2016}
Seth Flaxman, Sharad Goel, and Justin~M. Rao.
\newblock Filter {Bubbles}, {Echo} {Chambers}, and {Online} {News}
  {Consumption}.
\newblock {\em Public Opinion Quarterly}, 80:298--320, 2016.

\bibitem[FYM18]{fownes_twitter_2018}
Jennifer~R. Fownes, Chao Yu, and Drew~B. Margolin.
\newblock Twitter and climate change.
\newblock {\em Sociology Compass}, 12(6):e12587, 2018.
\newblock \_eprint: https://onlinelibrary.wiley.com/doi/pdf/10.1111/soc4.12587.

\bibitem[Gar09]{garrett_echo_2009}
R.~Kelly Garrett.
\newblock Echo chambers online?: {Politically} motivated selective exposure
  among {Internet} news users1.
\newblock {\em Journal of Computer-Mediated Communication}, 14(2):265--285,
  January 2009.

\bibitem[GGR10]{golbeck_twitter_2010}
Jennifer Golbeck, Justin~M. Grimes, and Anthony Rogers.
\newblock Twitter use by the {U}.{S}. {Congress}.
\newblock {\em Journal of the American Society for Information Science and
  Technology}, 61(8):1612--1621, 2010.
\newblock \_eprint: https://onlinelibrary.wiley.com/doi/pdf/10.1002/asi.21344.

\bibitem[HHvR{\etalchar{+}}07]{hallahan_defining_2007}
Kirk Hallahan, Derina Holtzhausen, Betteke van Ruler, Dejan Verčič, and
  Krishnamurthy Sriramesh.
\newblock Defining {Strategic} {Communication}.
\newblock {\em International Journal of Strategic Communication}, 1(1):3--35,
  March 2007.

\bibitem[Kee99]{keeley_conspiracy_1999}
Brian~L. Keeley.
\newblock Of {Conspiracy} {Theories}.
\newblock {\em The Journal of Philosophy}, 96(3):109, January 1999.
\newblock Publisher: Philosophy Documentation Center.

\bibitem[KKM13]{kleemann_interpretative_2013}
Frank Kleemann, Uwe Krähnke, and Ingo Matuschek.
\newblock {\em Interpretative {Sozialforschung}: {Eine} {Einführung} in die
  {Praxis} des {Interpretierens}}.
\newblock Springer VS, Wiesbaden, January 2013.

\bibitem[LSL13]{lamarre_tweeting_2013}
Heather~L. LaMarre and Yoshikazu Suzuki-Lambrecht.
\newblock Tweeting democracy? {Examining} {Twitter} as an online public
  relations strategy for congressional campaigns’.
\newblock {\em Public Relations Review}, 39(4):360--368, November 2013.

\bibitem[Lum97]{christoph_lumer_habermas_1997}
Christoph Lumer.
\newblock Habermas’ {Diskursethik}.
\newblock {\em Zeitschrift für philosophische Forschung}, 51(1):42--64,
  January 1997.
\newblock Publisher: Vittorio Klostermann GmbH.

\bibitem[Mer12]{mergel_connecting_2012}
Ines Mergel.
\newblock “{Connecting} to {Congress}”: {The} use of {Twitter} by {Members}
  of {Congress}.
\newblock {\em Zeitschrift für Politikberatung}, 5(3):108--114, 2012.

\bibitem[Ngu20]{nguyen_echo_2020}
C.~Thi Nguyen.
\newblock Echo {Chambers} and {Epistemic} {Bubbles}.
\newblock {\em Episteme}, 17(2):141--161, January 2020.

\bibitem[RJ20]{romer_conspiracy_2020}
Daniel Romer and Kathleen~Hall Jamieson.
\newblock Conspiracy theories as barriers to controlling the spread of
  {COVID}-19 in the {U}.{S}.
\newblock {\em Social science \& medicine (1982)}, 263:113356, January 2020.

\bibitem[Sco88]{scott_social_1988}
John Scott.
\newblock Social {Network} {Analysis}.
\newblock {\em Sociology}, 22(1):109--127, January 1988.

\bibitem[SP22]{silva_politicians_2022}
Bruno~Castanho Silva and Sven-Oliver Proksch.
\newblock Politicians unleashed? {Political} communication on {Twitter} and in
  parliament in {Western} {Europe}.
\newblock {\em Political Science Research and Methods}, 10(4):776--792, October
  2022.
\newblock Publisher: Cambridge University Press.

\bibitem[Swe11]{sweller_cognitive_2011}
John Sweller.
\newblock Cognitive {Load} {Theory}.
\newblock In Jose Mestre, editor, {\em Cognition in education}, volume~55 of
  {\em The psychology of learning and motivation}, pages 37--76. Elsevier,
  Amsterdam, January 2011.

\bibitem[Tho08]{thompson_deliberative_2008}
Dennis~F. Thompson.
\newblock Deliberative {Democratic} {Theory} and {Empirical} {Political}
  {Science}.
\newblock {\em Annual Review of Political Science}, 11(1):497--520, January
  2008.

\bibitem[Tor05]{strategic_turn}
Simon~Moberg Torp.
\newblock The strategic turn in communication science.
\newblock In Derina Holtzhausen and Ansgar Zerfass, editors, {\em The Routlet
  Handbook of Strategic Communication}. Routledge, 2005.

\bibitem[WD17]{wenninger_social_2017}
Andreas Wenninger and {Deutsche Akademie der Technikwissenschaften}.
\newblock {\em Social {Media} und digitale {Wissenschaftskommunikation}:
  {Analyse} und {Empfehlungen} zum {Umgang} mit {Chancen} und {Risiken} in der
  {Demokratie}}.
\newblock 2017.
\newblock OCLC: 1011375637.

\bibitem[YAM18]{brita_ytre-arne_approximately_2018}
Brita Ytre-Arne and H.~Moe.
\newblock Approximately {Informed}, {Occasionally} {Monitorial}?
  {Reconsidering} {Normative} {Citizen} {Ideals}.
\newblock {\em The International Journal of Press/Politics}, 23:227--246,
  January 2018.

\bibitem[ZBTM{\etalchar{+}}16]{filter_bubbles}
Frederik~J. Zuiderveen~Borgesius, Damian Trilling, Judith Möller, Balázs
  Bodó, Claes~H. Vreese, and Natali Helberger.
\newblock Should we worry about filter bubbles?
\newblock {\em Internet Policy Review}, 5(1), January 2016.

\end{thebibliography}

\end{document}